\documentclass[tightenlines,superscriptaddress,eqsecnum,floats,nofootinbib,showpacs]
{revtex4-1}

\usepackage{amssymb}
\usepackage{stmaryrd}
\usepackage{amsmath}
\usepackage{amsfonts}
\usepackage{mathrsfs}
\usepackage{CJK}
\usepackage{amsmath,amssymb,amsfonts}
\usepackage{graphicx}

\def\be{\begin{equation}}
\def\ee{\end{equation}}
\def\ba{\begin{eqnarray}}
\def\ea{\end{eqnarray}}
\def\nn{\nonumber}

\newcommand{\mubar}{{\bar \mu}} 
\newcommand{\abs}[1]{{\left|{#1}\right|}} 
\newcommand{\ket}[1]{\vert{#1}\rangle} 
\newcommand{\bra}[1]{\langle{#1}\vert} 

\newcommand{\sgn}{\mathrm{sgn}} 
\newcommand{\Tr}{\mathrm{Tr}} 
\newcommand{\grav}{\mathrm{gr}} 
\newcommand{\sca}{\mathrm{sc}} 
\newcommand{\kin}{\mathrm{kin}} 
\newcommand{\hil}{\mathcal{H}} 

\begin{document}


\title{ Higher dimensional Loop Quantum Cosmology}

\author{Xiangdong Zhang\footnote{xiangdong.zhang@gravity.fau.de}}
\affiliation{Department of Physics, South China University of
Technology, Guangzhou 510641, China}
\affiliation{Institute for Quantum Gravity, University of Erlangen-N{\"u}rnberg, Staudtstra{\ss}e 7 / B2, 91058 Erlangen, Germany}

\begin{abstract}

Loop quantum cosmology(LQC) is the symmetric model of loop quantum gravity. In this paper, we generalize the structure of loop quantum cosmology to the theories with arbitrary spacetime dimensions. The isotropic and homogenous cosmological model in n+1 dimensions is quantized by the loop quantization method.  Interestingly, we find that the underlying quantum theories are divided into two qualitatively different sectors according to spacetime dimensions. The effective Hamiltonian and modified dynamical equations of n+1 dimensional LQC are obtained. Moreover, our results indicate that the classical big bang singularity is resolved in arbitrary spacetime dimensions by a quantum bounce. We also briefly discuss the similarities and differences between the n+1 dimensional model and the 3+1 dimensional one. Our model serves as a first example of higher dimensional loop quantum cosmology and offers possibility to investigate quantum gravity effects in higher dimensional cosmology.
\end{abstract}


\maketitle

\section{Introduction}\label{sec:introduction}

Higher dimensional spacetime are a subject of great interest in the grand unify theories. Historically, the first higher dimensional theories is the famous Kaluza-Klein theory which trying to unify the 4 dimensional general relativity(GR) and Maxwell theory\cite{Appelguist}. Recent theoretical developments reveal that higher dimensions are perfered by many theories, such as the string/M theories\cite{JP98}, AdS/CFT correspondence\cite{JM03}, Brane world scenario\cite{RS99a,RS99b} and so on.  In the past decades, many aspects of these higher dimensional theories have been extensively studied. Particularly on the issues related to Black holes and cosmology. In fact, higher dimensional cosmology receives increasing attentions, and becomes a rather active field with fruitful results. For instance, some of the higher dimensional cosmological models can naturally explain the accelerated expansions of Universe\cite{NM02,Qiang}.

Loop quantum gravity(LQG) is a quantum gravity theory trying to quantize GR with the nonperturbative technicals\cite{Ro04,Th07,As04,Ma07}. Many issues of LQG have been explored in the past thirty years. Among these issues, loop quantum cosmology(LQC), which is the cosmological application of LQG receives particularly interest. Recently, LQC becomes one of  the most thriving and fruitful directions of LQG\cite{LQC5,Boj,Ash-view,AS11,BCM}. One of the most attractive features of this theory is that LQC is a singularity free theory. In LQC, the cosmological singularity which is inevitable in classical GR is naturally replaced by a quantum bounce\cite{APS3,ACS}. Although LQC a fruitful theory, nowadays all the discussions are still limited to the four spacetime dimensions. Recently, LQC has been generalized to the 2+1 dimensional case\cite{Zhang14}. Hence one is naturally to ask is it possible to generalize the structure of LQC to the higher spacetime dimensions?

 However, this is not an easy task, essentially because LQG is a quantization scheme based on the connection dynamics, while the $SU(2)$ connection dynamics is only well defined in three and four dimensions. And thus can not been generalized to the higher dimensional gravity theories. Fortunately, this difficulty has been overcame by Thiemann et. al. in the series of paper\cite{BTTa,BTTb,BTTc,BTTd}. The main idea of \cite{BTTa} is that in n+1 dimensional GR, in order to obtain a well defined connection dynamics, one should adapt $SO(n+1)$ connections $A_a^{IJ}$ rather than the speculated $SO(n)$ connections. With this higher dimensional connection dynamics in hand, Thiemann et. al successfully generalize the LQG to arbitrary spacetime dimensions.  Since the $2+1$ dimensional case is already studied in \cite{Zhang14}. The purpose of this paper is to investigate the issue of the n+1 dimensional LQC, with $n\geq3$ under this generalized LQG framework.

This paper is organized as follows: After a brief introduction. In Section \ref{section2}, we first review the classical connection formalism of n+1 dimensional LQG, and then we use it to derive the cosmological Hamiltonian through the symmetric reduction procedure. In section \ref{section3} we give a detailed construction of the quantum theory of $n+1$ dimensional LQC and derive the difference equation which represents the dynamical evolution of the $n+1$ dimensional Universe. Then we briefly discuss the singularity resolution issue in the \ref{section4}. The effective Hamiltonian as well as the modified effective dynamical equations are obtained in section \ref{section5} and \ref{section6} respectively. Some conclusions are given in the last section.

\section{classical theory}\label{section2}

To make this paper self-contained and also convenient for the readers, we first review some basic elements of classical $n+1$ dimensional gravity concerned in this paper. The connection dynamics of n+1 dimensional gravity with a guage group $SO(n+1)$ or $SO(1, n)$  is obtained in \cite{BTTa}.  The Ashtekar formalism of n+1 dimensional gravity constitutes a $SO(1,n)$ connections $A^{IJ}_a$ and group value densitized vector $\pi^b_{IJ}$ defined on an oriented n dimensional manifold $S$, where $a,b=1,2.....£¬n$ is the spatial indices and $I,J=1,2,3\dots n$ denotes $SO(1,n)$ group indices. The commutation relation for the canonical conjugate pairs satisfies \ba
\{A_{aIJ}(x),\pi^{bKL}(y)\}=2\kappa\gamma\delta^K_{[I}\delta^L_{J]}\delta_a^b\delta(x,y)
\ea where $\kappa=8\pi G$ and $\gamma$ is a nonzero real number. Here the $\pi^{bKL}$ satisfy the ``Simplicity constraint" \cite{BTTa} and can be written as $\pi^{bKL}=2n^{[K}E^{b|L]}=2\sqrt{h}h^{ab}n^{[K}e^{L]}_a$, where the spatial metric reads $h_{ab}=e_a^ie_{bi}$, $n^K$ is a normal which satisfy $e_a^Kn_K=0$ and $n^Kn_K=-1$. Moreover, the densitized vector $E^a_I$ satisfies $hh^{ab}=E^a_IE^{bI}$, where $h$ is the determinant of the spatial metric $h_{ab}$. $A_a^{IJ}$ is a $SO(1,n)$ connection defined as $A_a^{IJ}=\Gamma_a^{IJ}+\gamma K_a^{IJ}$, here $\Gamma_a^{IJ}$ and $K^{IJ}_{a}$ are n dimensional spin connection and extrinsic curvature respectively. Besides the Simplicity constraint, the n+1 dimensional gravity has three constraints similar with 3+1 dimensional general relativity\cite{BTTa,BTTc} \ba
G^{IJ}&:=&\mathcal {D}_{a}\pi ^{aIJ}:=\partial _{a}\pi ^{aIJ}+2A_{a}^{[I}{}_{K}\pi ^{a|K|J]}, \label {Gauss}\\
V_a&=&\frac {1}{2\gamma}F_{abIJ}\pi ^{bIJ},\label {vectornew}\\
H_{gr}&=&\frac{1}{2\kappa\sqrt{h}}\Big(F_{abIJ}\pi^{aIK}\pi^b{}_{K}{}^J+4\bar{D}^{aIJ}_T(F^{-1})_{aIJ, bKL}\bar{D}^{bKL}_T
-2(1+\gamma^2)K_{aI}K_{bJ}E^{a[I}E^{b|J]}\Big) \label{hamiltong}
\ea where $F_{abIJ}\equiv2\partial_{[a}A_{b]IJ}+2A_{a[I|K}A_{b}{}^{K}{}_{|J]}$ is the curvature of connection $A_{aIJ}$, and $\bar{D}^{aIJ}_T=\frac{\gamma}{4}F^{aIJ,bKL}\bar{K}^T_{bKL}$ with $\bar{K}^T_{bKL}$ being the transverse and traceless part of extrinsic curvature $K_{bKL}$.
Moreover, we have $[F\cdot F^{-1}]^{aIJ}_{bKL}=\delta^a_b\bar{\eta}^I_{[K}\bar{\eta}^J_{L]}$ with $\bar{\eta}^I_J=\delta^I_J-n^In_J$.

Now let us consider the $n+1$ dimensional isotropic and homogenous $k=0$ Universe. Its line element is described by the n+1 dimensional Friedmann-Robertson-Walker (FRW) metric
\ba
ds^2=-N^2dt^2+a^2(t)d\Omega^2 \label{lineelement}\nn
\ea
where $a$ is the scale factor and $d\Omega^2$ is the n dimensional sphere. We choose a fiducial
Euclidean metric  $ {}^oq_{ab}$ on the spatial slice of the isotropic observers and introduce a pair of fiducial orthnormal
basis as $({}^oe^a_I , {}^o\omega^I_a)$
such that $ {}^oq_{ab}={}^o\omega^I_a{}^o\omega^I_b$.
The physical spatial metric is related to the fiducial one by $ q_{ab}=a^2 {}^oq_{ab}$. Then the densitized  vector can be expressed as $E^a_I=pV_0^{-\frac{n-1}{n}}\sqrt{{}^0q}{}^oe^a_I
$, thus the $\pi^{aIJ}$ and spin connection $A_b^{IJ}$ respectively reduce to \ba
\pi^{aIJ}&=&2pV_0^{-\frac{n-1}{n}}\sqrt{{}^0q}{}^on^{[I}{}^oe^{a|J]}=pV_0^{-\frac{n-1}{n}}\sqrt{{}^0q}{}^o\pi^{aIJ}\\
A_b^{IJ}&=&2cV_0^{-\frac{1}{n}}{}^on^{[I}{}^o\omega^{J]}_b=cV_0^{-\frac{1}{n}}{}^{o}\Omega_a^{IJ}.
\ea In the following, for simplicity, we will fix the fiducial volume $V_0=1$. By using the classical expression $\pi^{aIJ}$, $A_b^{IJ}$ and comological line elements (\ref{lineelement}), one can easily yields \ba
p=a^{n-1}, \quad\quad c=\gamma\dot{a}
\ea
These canonical variables satisfy the commutation relation as follows\ba
\{c,p\}=\frac{\kappa\gamma}{n}\label{cp}
\ea
For our cosmological case, The Gaussian and diffeomophism constraints are satisfied automatically. For the Hamiltonian constraint, we first note that in our cosmological situation, the extrinsic curvature only has the diagonal part. Hence the transverse traceless part of extrinsic curvature $\bar{K}^T_{bKL}$ is identical to zero. Therefore the second term of the Hamiltonian constraint is vanishing. Moreover, the spin connection $\Gamma$ is also zero for our homogenous and isotropic Universe. Thus a simple straightforward calculation shows $KKEE$ term proportional to $F\pi\pi$ term. Combining all the above ingredients together, the Hamiltonian constraint (\ref{hamiltong}) reduces to \ba
H_{gr}&=&-\frac{1}{2\kappa\gamma^2}F_{abIJ}\frac{\pi^{aIK}\pi^b{}_{K}{}^J}{\sqrt{h}}
\ea Now, as in the 3+1 dimensional LQC, we also consider a minimally coupled massless scalar field $\phi$ as our matter field. The total Hamiltonian now reads
\ba
H_{Total}&=&-\frac{1}{2\kappa\gamma^2}F_{abIJ}\frac{\pi^{aIK}\pi^b{}_{K}{}^J}{\sqrt{h}}+\frac{p_\phi^2}{2\sqrt{h}}
\ea where the $p_\phi$ by definition is the conjugate momentum of massless scalar field $\phi$. The Poisson bracket between scalar field $\phi$ and conjugate momentum $p_\phi$ reads $\{\phi,p_\phi\}=1$. In the cosmological model we consider in this paper, this Hamiltonian therefore reduces to\ba
H_{Total}&=&-\frac{n(n-1)}{2\kappa\gamma^2}c^2p^{\frac{n-2}{n-1}}+\frac{p_\phi^2}{2p^{\frac{n}{n-1}}}\label{hamiltonht}
\ea Since at the classical level, this $SO(1,n)$ connection dynamics formalism is equivalent to the n+1 dimensional Arnowitt-Deser-Misner(ADM) formalism\cite{BTTa}. While in the cosmological situation, the ADM formalism will led to the classical Friedmann equation. Thus as a consistent check of our symmetric reduction procedure, we need to reproduce n+1 dimensional Friedmann equation with our Hamiltonian (\ref{hamiltonht}) and commutation relation (\ref{cp}).  To this aim, we calculate the equation of motion for $p$ which reads \ba
\dot{p}=\{p,H_{Total}\}=\frac{n-1}{\gamma}cp^{\frac{n-2}{n-1}}
\ea By using the Hamiltonian constraint we successfully reproduce the classical $n+1$ dimensional Friedmann equation \ba
H^2&=&\left(\frac{\dot{p}}{(n-1)p}\right)^2=\frac{1}{\gamma^2 p^2}c^2p^{\frac{2(n-2)}{n-1}}\nn\\
&=&\frac{2\kappa}{n(n-1)}\frac{p_\phi^2}{2p^{\frac{2n}{n-1}}}=\frac{2\kappa}{n(n-1)}\rho\label{Friedmann}
\ea where \ba
H&=&\frac{\dot{a}}{a}\nn\\
\rho&=&\frac{p_\phi^2}{2V^2}=\frac{p_\phi^2}{2p^{\frac{2n}{n-1}}}.
\ea are the Hubble parameter and the matter density in n+1 dimensions respectively. Moreover, another dynamical equation, namely the so-called Raychauduri equation which evolves second order time derivative of scale factor $a$ can be obtained by combining the continuity equation in $n+1$ dimensions $
\dot{\rho}+nH(\rho+p)=0$ with the Friedmann equation (\ref{Friedmann}) \ba
\frac{\ddot{a}}{a}=\frac{2\kappa}{n(n-1)}\rho-\frac{\kappa}{n-1}(\rho+p)
\ea

\section{Quantum theory}\label{section3}
Now we come to the issue of quantize the cosmological model, we first need to construct the
quantum kinematical Hilbert space of n+1 dimensional cosmology by mimicking the 3+1 dimensional loop
quantum cosmology. This quantum kinematical Hilbert spaces are constituted the so-called polymer-like quantization for geometric part while Schrodinger representation is adopted for the massless scalar field part. The resulted
kinematical Hilbert space for the geometry part reads
$\mathcal{H}_{\kin}^{\grav}\equiv L^2(R_{Bohr},d\mu_{H})$, where
$R_{Bohr}$ and $d\mu_{H}$ are respectively the Bohr
compactification of the real line $R$ and the corresponding Haar measure on it
\cite{LQC5}. On the other hand, follow the standard treatment of LQC, we choose
Schrodinger representation for the massless scalar field \cite{AS11}. Thus
the kinematical Hilbert space for the matter field part is defined as
in usual way as
$\mathcal{H}_{\kin}^{\sca}\equiv L^2(R,d\mu)$. Hence the whole Hilbert
space of n+1 dimensional loop quantum cosmology take the form of a direct product, $\hil_\kin:=\hil^\grav_\kin\otimes
\hil^\sca_\kin$. Now let $\ket{\mu}$ be the eigenstates of
 $\hat{p}$ in the kinematical Hilbert space $\mathcal{H}_{\kin}^{\grav}$ such that $
\hat{p}\ket{\mu}=\frac{4\pi G\gamma\hbar\mu}{n}\ket{\mu}=\frac{\hbar\kappa\gamma}{2n}\mu\ket{\mu}$.
These eigenstates $\ket{\mu_i}$ obey the orthonormal condition $
\bra{\mu_i}{\mu_j}\rangle=\delta_{\mu_i,\mu_j}$
with $\delta_{\mu_i,\mu_j}$ being the Kronecker delta function rather than the Dirac delta function. In n+1 dimensional quantum gravity, the $n-1$ dimensional area operator is quantized just like their counterparts in 3+1 dimensions, the discrete spectrum of this $n-1$ dimensional operator reads\cite{BTTc} \ba
\Delta_n=\kappa\hbar\gamma\sum_I\sqrt{I(I+n-1)}=8\pi\gamma(\ell_{\textrm{p}})^{n-1}\sum_I\sqrt{I(I+n-1)}
\ea where the $I$ is integers and $\ell_{\textrm{p}}=\sqrt[n-1]{G\hbar}$ being the Planck length. The interpretation of $I$ is that for every edge, we can associate a simple representations of $SO(n+1)$ which is labeled by its corresponding highest weight $\Lambda=(I,0,0,...)$ with $I$ being an integer. This equation tells us the existence of minimal area gap, which is given by \ba
\Delta_n=\sqrt{n}k\hbar\gamma\equiv 8\sqrt{n}\pi\gamma(\ell_{\textrm{p}})^{n-1}\label{minimumarea}
\ea Note that the quantization of area refers to
physical geometries in 3+1 dimensional LQC\cite{APS3}, we generalize this argument to our n+1 dimensional LQC. We take the n-1 dimensional cube, every vertex of the cube has n-1 edges, and the holonomy loop $\Box_{ij}$ constitutes by its arbitrary two edges from one vertex. Now we should shrink the holonomy loop $\Box_{ij}$ till
the n-1 dimensional area of the cube, which is measured by the physical metric
$q_{ab}$, reaches the value of minimal n-1 dimensional area $\Delta_n$. Since the physical n-1 dimensional area of
the elementary cell is $\abs{p}$ and each side of $\Box_{ij}$ is $\lambda$
times the edge of the elementary cell, in order to compare with 3+1 dimensions, we also use a specific function $\mubar(p)$ to denote $\lambda$, and similar to that in \cite{APS3}, we have \ba
\mubar^{n-1}(p)\abs{p}=\Delta_n\equiv8\sqrt{n}\pi\gamma(\ell_{\textrm{p}})^{n-1}
\ea
It is easy to see when $n=3$, the above formulation goes back to the famous $\mubar$ scheme in 3+1 dimensions. For the convenience of studying quantum dynamics, we define the following new
variables
\ba v:=\frac{2(n-1)\Delta_n}{\hbar\kappa\gamma}\mubar^{-n},\quad b:=\mubar c, \nn\ea
where
$\mubar=\left(\frac{\Delta_n}{|p|}\right)^{\frac{1}{n-1}}$ with
$\Delta_n$ being a minimum
nonzero eigenvalue of the n dimensional area operator \cite{Ash-view}. It is easy to verify that these new variables satisfy the commutation relation
$\{b,v\}=\frac{2}{\hbar}$.
It turns out that the eigenstates of
 $\hat{v}$ also constitute an orthonormal basis in the kinematical Hilbert space $\mathcal{H}_{\kin}^{\grav}$.
We denote
$\ket{\phi,v}$ as the generalized orthonormal basis for the whole kinematical Hilbert space
$\hil_\kin$. For simplicity, in the following, the $\ket{\phi,v}$ will be abbreviated as $\ket{v}$.

The action of volume operator $\hat{V}$ on this basis $\ket{v}$ reads \ba
\hat{V}\ket{v}=\frac{\hbar\kappa\gamma(\Delta_n)^\frac{1}{n-1}}{2(n-1)}\abs{v}\ket{v}=\frac{(\Delta_n)^\frac{n}{n-1}}{4(n-1)\sqrt{3}}\abs{v}\ket{v}
=\frac{4\pi\gamma(\Delta_n)^\frac{1}{n-1}}{(n-1)}\abs{v}\ell_{\textrm{p}}^{n-1}\ket{v}
\ea

The Hamiltonian constraint need to be reformulated in terms of these $(b,v)$ variables as \ba
H_{T}&=&-\frac{n(n-1)}{2\kappa\gamma^2}c^2p^{\frac{n-2}{n-1}}+\frac{p_\phi^2}{2p^{\frac{n}{n-1}}}\nn\\
&=&-\frac{n\hbar}{4\gamma(\Delta_n)^{\frac{1}{n-1}}}b^2\abs{v}+\left(\frac{2(n-1)}{\hbar\kappa\gamma (\Delta_n)^{\frac{1}{n-1}}}\right)\frac{p_\phi^2}{2\abs{v}}\label{shcm}
\ea

Note that we adapt polymer representation for geometric part, in the quantum theory, the connection should be replaced by the well defined holonomy operator.
In our cosmological setting the holonomy reads\ba
h^{\mubar}_{IJ}=\cos(\frac{\mubar c}{2})+2\tau_{IJ}\sin(\frac{\mubar c}{2})
\ea where $\tau_{IJ}=-\frac{i}{4}[\gamma_I,\gamma_J]$ with $\gamma_I$ being the gamma matrices constitutes a representation of $SO(n+1)$ . On the other hand, the curvature can be expressed through holonomy as
\ba
F_{abIJ}&=&-2\lim_{Ar_\Box\rightarrow 0}\Tr{\left(\frac{(h^{\mubar}_{\Box})_{KL,MN}-1}{\mubar^2}{}^o\Omega_a^{KL}{}^o\Omega_b^{MN}\tau_{IJ}\right)}\nn\\
&=&2\frac{\sin^2{(\mubar c)}}{\mubar^2}{}^o\Omega_{aK}^{[I}{}^o\Omega_b^{K|J]}\label{curvature}
\ea where we consider a square $\Box$ and \ba
(h^{\mubar}_{\Box})_{IJ,KL}=h_{IJ}h_{KL}h_{IJ}^{-1}h_{KL}^{-1}
\ea denotes the holonomy along a closed loop $\Box$. Every
edge of the square has length $\lambda (V_0)^{\frac1n}$
with respect to the fiducial
metric and the $Ar_\Box$ denotes the area of the square.

Now our task is to implement the Hamiltonian constraint at the quantum level. To this purpose, we first need to rewrite the Hamiltonian constraint with a suitable manner. This is inevitable because the expression of classical Hamiltonian constraint involves inverse of the determinate of n-metric and thus can not be promoted as a well defined operator on the kinematical Hilbert space. In 3+1 dimensional case, this difficulty can be overcome by using the well known classical identity $\frac12\epsilon^{ijk}\frac{\epsilon_{abc}E^b_jE^c_k}{\sqrt{q}}=\frac{1}{\kappa\gamma}\{A^i_a,V\}$\cite{Th07}. However, this expression need to be generalized in n+1 dimensional case. Interestingly, the treatment of the quantity $\frac{\pi^{[a|IK}\pi^{b]J}_K}{\sqrt{h}}$ can be divided into two different sectors according to spacetime dimensions, namely, even dimensional sector and odd dimensional sector\cite{BTTc}. First we note that\ba
\pi_{aIJ}(x):=-\frac{n-1}{2\kappa\gamma\sqrt{h}}\{A_{aIJ},V(x)\}
\ea now the quantity $\frac{\pi^{[a|IK}\pi^{b]J}_K}{\sqrt{h}}$ appearing in Hamiltonian constraint can be constructed with these basic building blocks.

\subsection{Even dimensional sector}
For the case of the spacetime dimensions $n+1$ are even. We let $s=\frac{(n-1)}{2}$, and note that we have the following classical identity \cite{BTTc}\ba
\frac{\pi^{[a|IK}\pi^{b]J}_K}{\sqrt{h}}&=&\frac{1}{4(n-2)!}\epsilon^{abca_1b_1\dots a_{s-1}b_{s-1}}\epsilon^{IJKLI_1J_1\dots I_{s-1}J_{s-1}}\nn\\
&=&\pi_{cKL}\pi_{a_1I_1K_1}\pi_{b_1J_1}^{K_1}\dots \pi_{a_{s-1}I_{s-1}K_{s-1}}\pi_{b_{s-1}J_{s-1}}^{K_{s-1}}\sqrt{h}^{n-2}\label{T1}
\ea
Since in the quantum theory, the connection should be replaced by the well defined holonomy operator, thus we can rewrite the Hamiltonian constraint as follows \ba
H_{gr}&&=-\frac{1}{2\kappa\gamma^2}\int d\Sigma F_{abIJ}\frac{\pi^{[a|IK}\pi^{b]J}_K}{\sqrt{h}}=-\frac{1}{8(n-2)!\kappa^{n-1}\gamma^{n}}\epsilon^{abca_1b_1\dots a_{s-1}b_{s-1}}\epsilon^{IJKLI_1J_1\dots I_{s-1}J_{s-1}}(n-1)^{n-2}\frac{1}{2^{n-2}}\nn\\
&&\int d\Sigma \left(F_{abIJ}\{A_{cKL},V\}\{A_{a_1I_1K_1},V\}\{A^{K_1}_{b_1J_1},V\}\dots \{A_{a_{s-1}I_{s-1}K_{s-1}},V\}\{A^{K_{s-1}}_{b_{s-1}J_{s-1}},V\}\right)\nn\\
\ea

At the quantum level, the connection is not a well defined operator, thus we replace it by holonomy. To this aim, first we observe that $\{A_{a}^{IJ},V\}\tau_{IJ}=\{c\tau_{IJ},V\}{}^o\Omega_a^{IJ}=-\frac{1}{\mubar}h_{IJ}\{h^{-1}_{IJ},V\}{}^o\Omega_a^{IJ}$. Moreover, Since we have the following identity \ba
\sqrt{{}^0q}=\det({}^o\Omega_a^{IJ})&=&\frac{1}{2n!}\epsilon^{abca_1b_1\dots a_{s-1}b_{s-1}}\epsilon^{IJKLI_1J_1\dots I_{s-1}J_{s-1}}\nn\\
&&\left({}^o\Omega_{aIM}{}^o\Omega^M_{bJ}{}^o\Omega_{cKL}{}^o\Omega_{a_1I_1K_1}{}^o\Omega^{K_1}_{b_1J_1}\dots {}^o\Omega_{a_{s-1}I_{s-1}K_{s-1}}{}^o\Omega^{K_{s-1}}_{b_{s-1}J_{s-1}}\right)
\ea according our convention, the spatial integral of above equation gives $\int d\Sigma \sqrt{{}^0q}=V_0=1$. Combining these facts with Eq.(\ref{curvature}) and use the commutator to replace the Poisson bracket, we obtain the exact expression of Hamiltonian constraint:\ba
\hat{H}_\grav&=&\frac{(-1)^{n-2}n(n-1)^{n-1}}{2(i\hbar)^{n-2}\kappa^{n-1}\gamma^{n}\mubar^n}\sin^2(\mubar c)\left(\sin(\frac{\mubar c}{2})\hat{V}\cos(\frac{\mubar c}{2})-\cos(\frac{\mubar c}{2})\hat{V}\sin(\frac{\mubar c}{2})\right)^{n-2}\nn\\
&=&\sin(\mubar c) \hat{F}\sin(\mubar c)\label{Hamiltonianoperator}
\ea where the action of $\hat{F}$ on a quantum state $\Psi(v)$ is defined by \ba
\hat{F}\Psi(v)=- \frac{n\hbar}{2^n\gamma(\Delta_n)^{\frac{1}{n-1}}}\abs{v}\left(\abs{v-1}-\abs{v+1}\right)^{n-2}\Psi(v)\equiv F(v)\Psi(v)
\ea  Interestingly, when $n=3$, the above Hamiltonian operator reads \ba
\hat{H}_\grav&=&\frac{6i}{\hbar\kappa^2\gamma^3\mubar^3}\sin^2(\mubar c)\left(\sin(\frac{\mubar c}{2})\hat{V}\cos(\frac{\mubar c}{2})-\cos(\frac{\mubar c}{2})\hat{V}\sin(\frac{\mubar c}{2})\right)
\ea which has exactly the same form with 3+1 dimensional LQC Hamiltonian operator\cite{APS3}. The action of Hamiltonian operator $\hat{H}_\grav$ on a quantum state $\Psi(v)\in\hil_\kin$ led to a similar difference equation as that in 3+1 dimensional LQC \ba
\hat{H}_\grav\Psi(v)=f_+(v)\Psi(v+4)+f_0(v)\Psi(v)+f_-(v)\Psi(v-4)\label{Hconstraint}
\ea where \ba
f_+(v)&=& -\frac14F(v+2)= \frac{n\hbar}{2^{n+2}\gamma(\Delta_n)^{\frac{1}{n-1}}}(\abs{v+2})\left(\abs{v+1}-\abs{v+3}\right)^{n-2}\nn\\
f_-(v)&=&-\frac14F(v-2)\nn\\
f_0(v)&=& \frac14F(v+2)+\frac14F(v-2)
\ea
Now we turn to the inverse volume operator which is appearing in the matter field part. As such, we first define the quantity $\abs{p}^{-1/2}$ in the following way \ba
\abs{p}^{-1/2}=\sgn(p)\frac{4}{\kappa\gamma\mubar}\Tr{\left(\sum_{IJ}\tau^{IJ}h_{IJ}\{h_{IJ}^{-1},V^{\frac{n-1}{2n}}\}\right)}\label{pbasic}
\ea Note that under the replacement $\{,\}\rightarrow\frac{1}{i\hbar}[,],  $ we have \ba
\Tr{\left(\sum_{IJ}\tau^{IJ}h_{IJ}[h_{IJ}^{-1},V^{\frac{n-1}{2n}}]\right)}=\frac n2\left(\sin(\frac{\mubar c}{2})V^{\frac{n-1}{2n}}\cos(\frac{\mubar c}{2})-\cos(\frac{\mubar c}{2})V^{\frac{n-1}{2n}}\sin(\frac{\mubar c}{2})\right)
\ea Since in the classical level we have $V^{-1}=\abs{p}^{-\frac{n}{n-1}}$,  Thus the action of inverse volume operator on a quantum state $\Psi(v)$ is just the suitable power of the Eq.(\ref{pbasic}) as follows \ba
\widehat{V^{-1}}\Psi(v)&=&\left(\frac{2n}{\kappa\gamma \hbar(\Delta_n)^{\frac{1}{n-1}}}\right)^{\frac{2n}{n-1}}\left(\frac{\kappa\gamma \hbar(\Delta_n)^{\frac{1}{n-1}}}{2(n-1)}\right)^{\frac{n+1}{n-1}}v^{\frac{2}{n-1}}\abs{\abs{v+1}^{\frac{n-1}{2n}}-\abs{v-1}^{\frac{n-1}{2n}}}^\frac{2n}{n-1}\Psi(v)\nn\\
&=&\frac{2(n-1)}{\kappa\gamma \hbar(\Delta_n)^{\frac{1}{n-1}}}\left(\frac{n}{n-1}\right)^{\frac{2n}{n-1}}v^{\frac{2}{n-1}}\abs{\abs{v+1}^{\frac{n-1}{2n}}-\abs{v-1}^{\frac{n-1}{2n}}}^\frac{2n}{n-1}\Psi(v)\nn\\
&:=&B(v)\Psi(v)
\ea In the semiclassical region, namely in the large $v$ region, the eigenvalue of the inverse volume operator $\widehat{V^{-1}}$ approaches to its classical value and turns out to be \ba
\left(\frac{2(n-1)}{\kappa\gamma \hbar(\Delta_n)^{\frac{1}{n-1}}}\right)\frac{1}{\abs{v}}
\ea
Collecting all the above ingredients, and note that $\hat{p}_\phi \Psi(v,\phi)=-i\hbar\frac{\partial \Psi(v,\phi)}{\partial\phi}$, we finally obtain the full quantum Hamiltonian constraint
\ba
\hbar^2\frac{B(v)}{2}\frac{\partial^2\Psi(v,\phi)}{\partial^2\phi}=\hat{H}_\grav\Psi(v)
\ea

\subsection{Odd dimensional sector}
For the case of the spacetime dimensions $n+1$ are odd. We let $s=\frac{(n-2)}{2}$, and note that we have the following classical identity \cite{BTTc}\ba
\frac{\pi^{[a|IK}\pi^{b]J}_K}{\sqrt{h}}&=&\frac{1}{2(n-2)!}\epsilon^{aba_1b_1\dots a_{s}b_{s}}\epsilon^{IJKI_1J_1\dots I_{s}J_{s}}\nn\\
&=&n_{K}\pi_{a_1I_1K_1}\pi_{b_1J_1}^{K_1}\dots \pi_{a_{s}I_{s}K_{s}}\pi_{b_{s}J_{s}}^{K_{s}}\sqrt{h}^{n-2}\label{T2}
\ea where the $n^I$ can be written in terms of $\pi_{aIJ}$ as \ba
n^I&=&\frac{1}{n!}\epsilon^{a_1b_1\dots a_{s+1}b_{s+1}}\epsilon^{II_1J_1\dots I_{s+1}J_{s+1}}\nn\\
&=&\pi_{a_1I_1K_1}\pi_{b_1J_1}^{K_1}\dots \pi_{a_{s+1}I_{s+1}K_{s+1}}\pi_{b_{s+1}J_{s+1}}^{K_{s+1}}\sqrt{h}^{n-1}
\ea
Thus we can rewrite the Hamiltonian constraint as follows \ba
H_{gr}&&=-\frac{1}{2\kappa\gamma^2}\int d\Sigma F_{abIJ}\frac{\pi^{[a|IK}\pi^{b]J}_K}{\sqrt{h}}=-\frac{1}{4(n-2)!\kappa^{n-1}\gamma^{n}}\epsilon^{aba_1b_1\dots a_{s}b_{s}}\epsilon^{IJKI_1J_1\dots I_{s}J_{s}}(n-1)^{n-2}\frac{1}{2^{n-2}}\nn\\
&&\int d\Sigma \left(F_{abIJ}n_K\{A_{a_1I_1K_1},V\}\{A^{K_1}_{b_1J_1},V\}\dots \{A_{a_sI_{s}K_{s}},V\}\{A^{K_s}_{b_sJ_{s}},V\}\right)\nn\\
\ea
Follow the recipe prescribed last subsection, by replacing the connection by holonomy and the Poisson bracket by the commutator we obtain the  quantum Hamiltonian constraint operator \ba
\hat{H}_\grav&=&\frac{n(n-1)^{4n-3}2^{2n-3}}{(2n-3)^{2n-2}(i\hbar)^{2n-2}\kappa^{2n-1}\gamma^{2n}\mubar^{2n}}\sin^2(\mubar c)\left(\sin(\frac{\mubar c}{2})V^{\frac{2n-3}{2n-2}}\cos(\frac{\mubar c}{2})-\cos(\frac{\mubar c}{2})V^{\frac{2n-3}{2n-2}}\sin(\frac{\mubar c}{2})\right)^{2n-2}\nn\\
&=&\sin(\mubar c) \hat{F}\sin(\mubar c)
\ea where the action of $\hat{F}$ on a quantum state $\Psi(v)$ is defined by \ba
\hat{F}\Psi(v)=- \frac{n\hbar(n-1)^{2n-2}}{4(2n-3)^{2n-2}\gamma(\Delta_n)^{\frac{1}{n-1}}}\abs{v}^2\left(\abs{v-1}^{\frac{2n-3}{2n-2}}-\abs{v+1}^{\frac{2n-3}{2n-2}}\right)^{2n-2}\Psi(v)\equiv F(v)\Psi(v)
\ea  This operator acts on a quantum state $\Psi(v)\in\hil_\kin$, gives a difference equation \ba
\hat{H}_\grav\Psi(v)=f_+(v)\Psi(v+4)+f_0(v)\Psi(v)+f_-(v)\Psi(v-4)\label{Hconstraint}
\ea where \ba
f_+(v)&=& -\frac14F(v+2)= \frac{n\hbar(n-1)^{2n-2}}{16(2n-3)^{2n-2}\gamma(\Delta_n)^{\frac{1}{n-1}}}\abs{v+2}^2\left(\abs{v+1}^{\frac{2n-3}{2n-2}}-\abs{v+3}^{\frac{2n-3}{2n-2}}\right)^{2n-2}\nn\\
f_-(v)&=&-\frac14F(v-2)\nn\\
f_0(v)&=& \frac14F(v+2)+\frac14F(v-2)
\ea The action of the inverse volume operator keeps the same form as in even dimensional case. Therefore, the full quantum Hamiltonian constraint also turns out to be the following
\ba
\hbar^2\frac{B(v)}{2}\frac{\partial^2\Psi(v,\phi)}{\partial^2\phi}=\hat{H}_\grav\Psi(v)
\ea

\section{singularity resolution}\label{section4}

Now we come to deal with the issue of the singularity resolution. In order to proceed, we take the same strategy which adopted in \cite{ACS}. To be more specifically, we first make some reasonable simplifications on our quantum Hamiltonian constraint equation such that the whole dynamical system becomes simlper and exactly solvable. Then the discussion of the issue of singularity resolution will be made within this exactly solvable formalism\cite{ACS}.  We make the following replacements as in\cite{ACS}:
\ba B(v)\longmapsto \left(\frac{2(n-1)}{\kappa\gamma \hbar(\Delta_n)^{\frac{1}{n-1}}}\right)\frac{1}{\abs{v}},\nn
\ea
and
\ba F(v)\longmapsto- \frac{n\hbar}{4\gamma(\Delta_n)^{\frac{1}{n-1}}}\abs{v}. \nn\ea
The validity of the first replacement amounts to
assuming $\mathcal {O}(\frac{1}{\abs{v}})\ll 1$, which also in turn implies the second
replacement.

In the corresponding
quantum theory, the
Hamiltonian constraint equation now reduce to \ba
\frac{\partial^2\Psi(v)}{\partial\phi^2}&=&-\hat{\Theta}\Psi(v)\nn\\
&=&\frac{n\kappa}{4(n-1)}v \sin(b)v \sin(b)\Psi(v)\nn\\
&=&\frac{n\kappa}{16(n-1)}v\left[(v+2)\Psi(v+4)-2v\Psi(v)+(v-2)\Psi(v-4)\right]\label{hamilton}\ea
here we denote quantum state
$\Psi(v)\equiv\Psi(v,\phi)$ for short.
The Eq.(\ref{hamilton}) gives rise a Klein-Gordon type equation. The physical state for the quantum dynamics of n+1 dimensional LQC thus is given by the ``positive frequency" square root of the Eq.(\ref{hamilton}) as\ba
\frac{\partial\Psi(v)}{\partial\phi}=i\sqrt{\Theta}\Psi(v).
\ea Note that here exists a superselection ambuguity, namely, for any real number $\epsilon\in[0,4)$ the states $\Psi(v)$ supported on points $v=4k+\epsilon$ with $k$ being an integer lead to the same dynamics. Thus as in\cite{Zhang14} we just fix $\epsilon=0$. Moreover, note that because the state $\ket{0}$ has zero norm thus it is excluded out of the physical Hilbert space. The physical inner product between two states reads \ba
<\Psi_1,\Psi_2>_{phy}:=\frac{1}{\pi}\sum_{v=4k}\frac{1}{\abs{v}}\bar{\Psi}_1(v)\Psi_2(v)\label{physicalinner1}
\ea Note that $(b,v)$ forms a canonical conjugate pair, thus the Fourier transforms $\Psi(b)$ has a support on the interval $(0,\pi)$. Therefore the Fourier transformation and its corresponding inverse transformation are defined respectively as\ba
\Psi(b):= \sum_{v=4k}e^{\frac{i}{2}vb}\Psi(v),\quad\quad
\Psi(v)=\frac{1}{\pi}\int_0^\pi  e^{-\frac{i}{2}vb}\Psi(b) db
\ea
Now we set $\chi(v)=\frac{1}{\pi v}\Psi(v)$, then the constraint Eq.(\ref{hamilton}) becomes a second-order differential equation \ba
\frac{\partial^2\chi(b)}{\partial^2\phi}=\frac{n\kappa}{n-1}\sin^2(b)\frac{\partial^2\chi(b)}{\partial^2 b}\label{chib}
\ea We define the following new variable $x$ to make this equation more simpler,  \ba
x=\sqrt{\frac{n-1}{n\kappa}}\ln\left(\tan(\frac{b}{2})\right)
\ea Then the constraint Eq. (\ref{chib}) becomes the standard Klein-Gordon type equation \ba
\frac{\partial^2\chi(b)}{\partial^2\phi}=\frac{\partial^2\chi(b)}{\partial^2 x}\label{KG}
\ea The physical Hilbert space is the span of positive frequency solutions to Eq. (\ref{KG}). This equation can be further simplified if we decompose the solution into left and right moving sectors as
\ba
\chi(x)=\chi_L(x_+)+\chi_R(x_-),
\ea here $x_{\pm}=\phi\pm x$. In addition, $\chi(x)$ has a following symmetry \ba
\chi(-x)=-\chi(x).
\ea This feature enables us to make a further decomposition \ba
\chi(x)=\frac{1}{\sqrt{2}}\left(F(x_+)-F(x_-)\right),
\ea where $F(x_{\mp})$ by definition are negative/positive frequency solutions to Eq.(\ref{KG}). The physical inner product (\ref{physicalinner1}) now reads \ba
<\chi_1,\chi_2>_{phy}:=i\int_{-\infty}^{\infty}dx\left[\left(\frac{\partial\bar{F}_1(x_+)}{\partial x}\right)F_2(x_+)-\left(\frac{\partial\bar{F}_1(x_-)}{\partial x}\right)F_2(x_-)\right]
\ea Now the expectation value of the volume operator can be calculated as follows \ba
\langle \hat{V}\rangle|_\phi&=&(\chi,\hat{V}|_\phi\chi)_{phy}=\frac{\hbar\kappa\gamma(\Delta_n)^{\frac{1}{n-1}}}{2(n-1)}(\chi,\abs{\hat{v}}\chi)_{phy}\nn\\
&=&i\frac{\hbar\kappa\gamma(\Delta_n)^{\frac{1}{n-1}}}{2(n-1)}\int_{-\infty}^{\infty}dx\left[\left(\frac{\partial\bar{F}(x_+)}{\partial x}\right)(\hat{v}F(x_+))-\left(\frac{\partial\bar{F}(x_-)}{\partial x}\right)(-\hat{v}F(x_-))\right]\nn\\
&=&\frac{\hbar\kappa\gamma(\Delta_n)^{\frac{1}{n-1}}}{(n-1)\sqrt{\beta}}\int^{\infty}_{-\infty}dx\abs{\frac{\partial F}{\partial x}}^2\cosh(\sqrt{\beta}(x-\phi))\nn\\
&=&V_+e^{\sqrt{\beta}\phi}+V_-e^{-\sqrt{\beta}\phi}\label{Vexpectation}
\ea where $\beta=\frac{n\kappa}{n-1}$ and \ba
V_{\pm}=\frac{\hbar\kappa\gamma(\Delta_n)^{\frac{1}{n-1}}}{(n-1)\sqrt{\beta}}\int^{\infty}_{-\infty}\abs{\frac{\partial F}{\partial x}}^2e^{\mp\sqrt{\beta}x} dx
\ea From Eq.(\ref{Vexpectation}), it is clear that the expectation value of $\hat{V}$ admits a nonzero minimum $
V_{min}=2\sqrt{V_+V_-}$. This implies that all states undergo a big bounce rather than experience a singularity which has zero expectation of volume operator.  To justify this conclusion, let us turn to matter density $\rho=\langle\rho|_{\phi_0}\rangle$. if our picture is right, this important physical observable should have a upper bound. The classical definition of matter density reads $\rho=\frac{p^2_\phi}{2V^2}$ and we can see that the matter density will to into infinity at the singularity point as the volume will going to zero. Thus as a comparison we calculate the expectation value of $\rho$, if the singularity is really resolved, the expectation value of matter density must have a upper bound. To this aim, we first need to know the matrix elements of the observable $\hat{p}_\phi$, which reads \ba
\frac{1}{\hbar} <F_1,\hat{p}_\phi F_2>_{phy}=\int^{\infty}_{-\infty}\left(\frac{\partial\bar{F}_1(x)}{\partial x}\right)\frac{\partial F_2(x)}{\partial x} dx
\ea   Now we use a fixed state  $\chi(x)=\frac{1}{\sqrt{2}}\left(F(x_+)-F(x_-)\right)$ to calucate the expectation value of matter denstiy at the moment of $\phi_0$ \ba
\langle\rho|_{\phi_0}\rangle&=&\frac{(\langle \hat{p}_\phi\rangle)^2}{2(\langle\hat{V}\rangle)^2}\nn\\
&=&\frac{(n-1)^2\beta\hbar^2}{2\hbar^2\kappa^2\gamma^2 (\Delta_n)^{\frac{2}{n-1}}}\frac{\left[\int^{\infty}_{-\infty} dx\abs{\frac{\partial F}{\partial x}}^2\right]^2}{\left[\int^{\infty}_{-\infty} dx\abs{\frac{\partial F}{\partial x}}^2\cosh(\sqrt{\beta}x)\right]^2}\nn\\
&\leq&\frac{n(n-1)}{2\kappa\gamma^2 (\Delta_n)^{\frac{2}{n-1}}}=\rho_c
\ea where we use the fact $\cosh(\sqrt{\beta}x)\geq1$ in the second line. An interesting fact is that, in section \ref{section5} we will found that this upper-bound of the expectation value of matter density coincides with the critical matter density which yields from the effective Friedmann equation.

\section{Effective Hamiltonian}\label{section5}

One of the most delicate and valuable issue is the effective description of LQC,
since it predicts the possible quantum gravity effects to low-energy
physics. Both the
canonical \cite{Taveras,DMY,YDM,Boj11} and path integral
perspective\cite{ACH102,QHM,QDM,QM1,QM2} of the effective Hamiltonian of LQC has been studied.

With the Hamiltonian constraint equation (\ref{hamilton}) in hand, we now derive the effective
Hamiltonian within the $n+1$ dimensional timeless path integral formalism.
In the timeless path integral formalism, the dynamics are encoded into the transition amplitude which equals to the physical inner product \cite{ACH102,QHM}, i.e.,
\begin{align}
A(v_f, \phi_f;~v_i,\phi_i)=\langle v_f,
\phi_f|v_i,\phi_i\rangle_{phy}=\lim\limits_{\alpha_o\rightarrow\infty}
\int_{-\alpha_o}^{\alpha_o}d\alpha\langle
v_f,\phi_f|e^{i\alpha\hat{C}}|v_i,\phi_i\rangle, \label{amplitude}
\end{align}
where the subscript of $i$ and $f$ represent for initial and final state, and $\hat{C}\equiv\hat{\Theta}+\hat{p}_{\phi}^2/\hbar^2$. As shown in \cite{QHM,QDM}, by inserting some suitable complete basis and do multiple group averaging, Eq(\ref{amplitude}) is equivalent to calculate
\begin{align}
\langle v_f, \phi_f|e^{i\sum\limits_{m=1}^I{\epsilon\alpha_m}\hat{C}}|v_i,\phi_i\rangle=\sum\limits_{v_{I-1},...v_1}\int d\phi_{I-1}...d\phi_1\prod\limits_{m=1}^I\langle \phi_m|\langle v_m|e^{i\epsilon\alpha_m\hat{C}}|v_{m-1}\rangle|\phi_{m-1}\rangle.
\label{insert basis}
\end{align}
Note that the action of the constraint equation has the Klein-Gordon type, and thus its action on the gravitational part and the scalar field part can be calculated separately. So we first calculate the matter part and get
\begin{align}
\langle{\phi_m}|e^{i\epsilon\alpha_m\frac{\widehat{p}^2_\phi}{\hbar^2}}|\phi_{m-1}\rangle
=&\int dp_{\phi_m}\langle{\phi_m}|p_{\phi_m}\rangle\langle p_{\phi_m}|e^{i\epsilon\alpha_m\frac{\widehat{p}^2_\phi}{\hbar^2}}|\phi_{m-1}\rangle\nonumber\\
=&\frac{1}{2\pi\hbar}\int dp_{\phi_m}e^{i\epsilon(\frac{p_{\phi_m}}{\hbar}\frac{\phi_m-\varphi_{m-1}}{\epsilon}
+\alpha_m\frac{{p}^2_{\phi_m}}{\hbar^2})}.
\label{material amplitude}
\end{align}
For the gravity part, we expand the exponential and neglect the higher order terms give us
\ba \int
d\phi_{m}\bra{\phi_m}\bra{v_m}e^{-i\epsilon\alpha_m\hat{\Theta}}\ket{v_{m-1}}\ket{\phi_{m-1}}=
\delta_{v_m,v_{m-1}}-i\epsilon\alpha_m\int
d\phi_{m}\bra{\phi_m}\bra{v_m}\hat{\Theta}\ket{v_{m-1}}\ket{\phi_{m-1}}+\mathcal
{O}(\epsilon^2). \label{GHamilton}\ea
By using Eq.(\ref{hamilton}), the matrix elements of $\bra{\phi_m}\bra{v_m}\hat{\Theta}\ket{v_{m-1}}\ket{\phi_{m-1}}$
can be evaluated as follows
\ba
&&2\pi\hbar\int d\phi_m\bra{\phi_m}\bra{v_m}\hat{\Theta}\ket{v_{m-1}}\ket{\phi_{m-1}}\nn\\
&=&\int d\phi_{m}
dp_{\phi_{m}}e^{i\epsilon(\frac{p_{\phi_{m}}}{\hbar}\frac{\phi_m-\phi_{m-1}}{\epsilon})}\frac{n\kappa}{16(n-1)}v_{n-1}
\frac{v_m+v_{m-1}}{2}
(\delta_{v_m,v_{m-1}+4}-2\delta_{v_mn,v_{m-1}}+\delta_{v_m,v_{m-1}-4}),\nn
\ea
By using the following identity
\ba
\delta_{v_m,v_{m-1}+4}-2\delta_{v_m,v_{m-1}}+\delta_{v_m,v_{m-1}-4}=\frac{4}{\pi}\int_0^\pi
db_me^{-ib_m(v_m-v_{m-1})/2}\sin^2(b_m),\nn
\ea
Eq.(\ref{GHamilton}) can be rewritten in a compact form
\ba
&&2\pi\hbar\int
d\phi_{m}\bra{\phi_m}\bra{v_m}e^{-i\epsilon\alpha_m\hat{\Theta}}\ket{v_{m-1}}\ket{\phi_{m-1}}\nn\\
&=&\int d\phi_{m}dp_{\phi_{m}}e^{i\epsilon(\frac{p_{\phi_{m}}}{\hbar}\frac{\phi_m-\phi_{m-1}}{\epsilon})}\frac{1}{\pi}\int_0^\pi
db_ne^{-ib_m(v_m-v_{m-1})/2}\left[1-i\alpha_m\epsilon\frac{n\kappa}{16(n-1)}v_{m-1}\frac{v_m+v_{m-1}}{2}4\sin^2b_m\right]. \nn\ea
Combining all the above ingredients together, the physical transition amplitude can be written as follows
\ba
&&A(v_f, \phi_f;~v_i,\phi_i)\nn\\
&=&\lim\limits_{I\rightarrow\infty}~\lim\limits_{\alpha_\emph{{Io}},...,\alpha_\emph{{1o}}\rightarrow\infty}
\left(\epsilon\prod\limits_{m=2}^I\frac{1}{2\alpha_\emph{{mo}}}\right)\int_{-\alpha_\emph{{Io}}}^{\alpha_\emph{{Io}}} d\alpha_I...\int_{-\alpha_\emph{{1o}}}^{\alpha_\emph{{1o}}} d\alpha_1\nonumber\\
&\times&\int_{-\infty}^{\infty}d\phi_{I-1}...d\phi_1\left(\frac{1}{2\pi\hbar}\right)^I\int_{-\infty}^{\infty}
dp_{\phi_{I}}...dp_{\phi_{1}}\sum\limits_{v_{I-1},...,v_1}~\left(\frac{1}{\pi}\right)^I\int^{\pi}_{0}db_I...db_1\nonumber\\
&\times&\prod\limits_{m=1}^{I}\exp{i\epsilon}\left[
\frac{p_{\phi_{m}}}{\hbar}\frac{\phi_m-\phi_{m-1}}{\epsilon}
-\frac{b_m}{2}\frac{v_m-v_{m-1}}{\epsilon}+\alpha_m \left(\frac{p_{\phi_m}^2}{\hbar^2}-\frac{n\kappa}{16(n-1)}v_{m-1}\frac{v_m+v_{m-1}}{2}4\sin^2b_m\right)\right].\nn
\ea Under the `continuum limit', the finally result of transition amplitude reads
\ba
&&A(v_f, \phi_f;~v_i,\phi_i)\nn\\
&=&c\int \mathcal{D}\alpha\int\mathcal{D}\phi\int\mathcal{D}p_{\phi}\int\mathcal{D}v\int\mathcal{D}b ~~\exp\left(
\frac{i}{\hbar}\int_0^1d\tau \left[p_\phi\dot\phi-\frac{\hbar b}{2}\dot{v}+{\hbar}{\alpha}\left(\frac{p_\phi^2}{\hbar^2}
-\frac{n\kappa}{4(n-1)}v^2\sin^2b\right)\right]\right),\nn
\ea
where $c$ is an overall constant which do not affect the dynamics.
Hence, the effective Hamiltonian constraint in our $n+1$ dimensional LQC model can be simply read-off as
\ba
C_{eff}&=&\frac{p_\phi^2}{\hbar^2}-\frac{n\kappa}{4(n-1)}v^2\sin^2b\nn\\
&=&\frac{p_\phi^2}{\hbar^2}-\frac{n(n-1)\Delta_n^2}{\hbar^2\kappa\gamma^2\mubar^{2n}}\sin^2(\mubar c)\label{Ceff}\ea
When we take the large scale limit which by definition is $\sin b\rightarrow b$ (or $\sin(\mubar c)\rightarrow \mubar c$ in (c,p) representation). We observed that the classical Hamiltonian constraint (\ref{shcm}) is recovered from Eq.(\ref{Ceff}) up to a inverse volume factor $\frac{1}{|V|}$. The reason for this is simply because the Hamiltonian constraint in the previous sections we are not using the proper time of isotropic observers. To considering this point, the factor $\frac{1}{|V|}$ then has to be multiplied to $C_{eff}$ to obtain the correct result. As a consequence, we finally yield the physical effective Hamiltonian
\ba
H_F=\frac{1}{\abs{v}}C_{eff}=-\frac{n\hbar}{4\gamma(\Delta_n)^{\frac{1}{n-1}}}\abs{v}\sin^2b+\frac{\hbar\kappa\gamma (\Delta_n)^{\frac{1}{n-1}}}{2(n-1)}\abs{v}\rho\label{physicalH},
\ea
where the $n+1$ dimensional matter density by definition is $\rho=\frac{2(n-1)^2p_\phi^2}{v^2\hbar^2\kappa^2\gamma^2(\Delta_n)^{\frac{2}{n-1}}}$.

\section{Effective equation}\label{section6}
Now we are ready to derive the physical evolution equation of the $n+1$ dimensional Unverse, the most important one is of course the modified Friedmann equation. To this aim, we combining the effective Hamiltonian constraint $H_F$ (\ref{physicalH}) with the symplectic structure of $n+1$ dimensional loop quantum cosmology, one can easily obtain equations of motion for
the volume $v$ and the scalar field $\phi$ respectively as
\ba
\dot{v}&=&\{v,H_F\}=\frac{n}{\gamma(\Delta_n)^{\frac{1}{n-1}}}|v|\sin
(b)\cos(b),\label{vdot}\\
\dot{\phi}&=&\{\phi,H_F\}=\frac{4(n-1)
p_\phi}{\hbar\kappa\gamma (\Delta_n)^{\frac{1}{n-1}}|v|}. \label{phidot1} \ea By using Eq.(\ref{vdot}), it is easy to
see that  \ba
H^2=\left(\frac{\dot{a}}{a}\right)^2=\left(\frac{\dot{v}}{nv}\right)^2=\frac{1}{\gamma^2(\Delta_n)^{\frac{2}{n-1}}}\sin^2
(b)\cos^2(b).\label{hubble}
\ea Since the right hand side of the above equation evolves $\sin^2
(b)\cos^2(b)$, thus has some relation with the effective Hamiltonian constraint (\ref{physicalH}). In fact, the the effective Hamiltonian constraint $H_F=0$
can be rewritten as the following compact form \ba
\sin^2b=\frac{2\kappa\gamma^2(\Delta_n)^{\frac{2}{n-1}}\rho}{n(n-1)}=\frac{\rho}{\rho_c}
\ea Here we define $\rho_c=\frac{n(n-1)}{2\kappa\gamma^2(\Delta_n)^{\frac{2}{n-1}}}$ as the $n+1$ dimensional critical matter density. As shown in section (\ref{section4}) this $\rho_c$ in fact is the upper bound of the matter density. With the help of this equation, the modified Friedmann equation reads \ba
H^2=\left(\frac{\dot{a}}{a}\right)^2=\frac{2\kappa}{n(n-1)}\rho\left(1-\frac{\rho}{\rho_c}\right)\label{effectiveH}
\ea From Eq. (\ref{effectiveH}), it is easy to see that when $\rho=\rho_c$, we have $\dot{v}=0$, which implies the existence of a quantum bounce
at that point. To justify this, we calculate the second derivative of $v$ at the point of $\rho=\rho_c$\ba
\ddot{v}|_{\rho=\rho_c}&=&\{\dot{v},H_F\}|_{\rho=\rho_c}=\frac{n^2}{\gamma^2(\Delta_n)^{\frac{2}{n-1}}}\abs{v}\neq0
\ea Obviously, this implies a quantum bounce
occurs at that point. Moreover£¬ combining the continuity equation(which is nothing but the equation of motion for scalar field $\phi$) in n+1 dimensions \ba
\dot{\rho}+nH(\rho+p)=0
\ea with Eq. (\ref{effectiveH}), we can easily obtain another dynamical equation of the n+1 dimensional Universe which is so-called Raychauduri equation with loop quantum correction\ba
\frac{\ddot{a}}{a}=\frac{2\kappa}{n(n-1)}\rho\left(1-\frac{\rho}{\rho_c}\right)-\frac{\kappa}{n-1}(\rho+p)\left(1-\frac{2\rho}{\rho_c}\right)
\ea

\section{conclusion}\label{section7}

In this paper, a detailed construction of the $n+1$ dimensional LQC is presented. We start from the classical connection dynamics of $n+1$ dimensional general relativity, and then using the nonperturbative loop quantization method, we find that the dynamical evolution of the $n+1$ dimensional Universe is fully determined by a difference equation. Interestingly, in the quantum theory, even dimensional sector and odd dimensional sector are exhibiting qualitative different features. In order to obtain the effective equations of n+1 dimensional LQC which contain quantum corrections to the classical equations. We then generalize the timeless path integral formalism of LQC to n+1 dimensional case and use it to derive the modified effective Hamiltonian of n+1 dimensional LQC. Based on this effective Hamiltonian, the Friedmann equation as well as the Raychauduri equation with loop quantum correction is obtained.  Our results indicate that the classical singularity is resolved by a quantum bounce in arbitrary spacetime dimensions. In addition, we find that the heuristic replacement $c\rightarrow\frac{\sin(\mubar c)}{\mubar}$ with $\mubar=\left(\frac{\Delta_n}{p}\right)^{\frac{1}{n-1}}$ works not only for the 3+1 dimensional case, but also for the more general dimensional case.

One of the interesting question of $n+1$ dimensional LQG with gauge group $SO(n+1)$ is that, when $n=3$, namely, in four dimensions, we have two LQG theories, One is usually $SU(2)$ LQG, while the other one is $SO(4)$ LQG. Does these two theories equivalent to each other? in this paper we show at least at LQC level, they are indeed equivalent to each other if we identify the critical matter density $\rho_c$ of these two theories. Since the critical matter density are closely related with the minimal area of the corresponding quantum gravity theories, and note that there has one parameter ambiguity $\gamma$ in the definition of minimal area in Eq.(\ref{minimumarea}). Thus at LQC level, the $SU(2)$ LQC and $SO(4)$ LQC in fact is the same theory with the difference choice of the immirzi parameter $\gamma$.

Our work offers possiblity to explore the issues of LQC with the spacetime dimension higher than four. In particular, nowadays, higher dimensional cosmology becomes a rather popular and active field. For example, by using the dimensional reduction method, the cosmic acceleration can be naturally explained by some five dimensional models \cite{Qiang}. Hence it is also very interesting to study these issues with our higher dimensional LQC, Moreover, results we developed in this paper lay a foundation for the phenomenological investigations to possible quantum gravity effects in higher dimensional cosmology.

Another interesting topic is to derive the LQC directly for the LQG. In 3+1 dimensional case, some interesting efforts have been made towards this important direction\cite{AC13a,AC13b,NB15}. It is quite interesting to discuss this topic in  our $n+1$ dimensional LQC setting, and we will leave all these interesting and delicate topics for future study.

\begin{acknowledgements}
The author would like to thank Prof. Thomas Thiemann and Dr. Muxin Han for helpful discussions. The author would also like to thank the financial supported by CSC-DAAD postdoctoral fellowship. This work is supported by NSFC with No.11305063  and
the Fundamental Research Funds for the Central University of China.

\end{acknowledgements}



\begin{thebibliography}{99}


\bibitem{Appelguist}T. Appelquist, A. Chodos, and P. G. O. Freund (ed.), Modern
Kaluza-Klein Theories, Frontiers in Physics Series
Vol.\textbf{65}(Addison-Wesley, Reading, MA, 1986).


\bibitem{JP98}J. Polchinski, {\it String Theory,} Vol 1 and Vol 2, (Cambridge University Press, 1998).

\bibitem{JM03}J. M. Maldacena, {\it TASI 2003 Lectures on AdS/CFT,} arXiv:hep-th/0309246.

\bibitem{RS99a}L. Randall and R. Sundrum, {\it Large Mass Hierarchy from a Small Extra Dimension,} Phys. Rev. Lett. \textbf{83}, 3370
(1999).

\bibitem{RS99b}L. Randall and R. Sundrum, {\it An Alternative to Compactification,} Phys. Rev. Lett. \textbf{83}, 4690
(1999).

\bibitem{NM02}N. Mohammedi, {\it Dynamical compactification, standard cosmology, and the accelerating universe,} Phys. Rev. D \textbf{65}, 104018(2002).

\bibitem{Qiang}L. Qiang, Y. Ma, M. Han and D. Yu, {\it 5-dimensional Brans-Dicke theory and cosmic acceleration,} Phys. Rev. D {\bf71},  061501(R)
(2005).

\bibitem{Ro04} C. Rovelli, {\it Quantum Gravity,} (Cambridge University Press, 2004).

\bibitem{Th07} T. Thiemann, {\it Modern Canonical Quantum General Relativity,} (Cambridge University
Press, 2007).


\bibitem{As04}A. Ashtekar and J. Lewandowski, {\it Background independent quantum gravity: A
status report,} Class. Quant. Grav. {\bf21}, R53 (2004).

\bibitem{Ma07} M. Han, W. Huang, and Y. Ma, {\it Fundamental structure of loop quantum gravity,}
 Int. J. Mod. Phys. D {\bf16}, 1397 ,(2007).


\bibitem{LQC5}
A. Ashtekar, M. Bojowald, and J. Lewandowski, {\it Mathematical structure of loop quantum cosmology,} Adv. Theor. Math.
Phys. \textbf{7}, 233 (2003).

\bibitem{Boj}
M. Bojowald, {\it Loop quantum cosmology,} Living Rev. Relativity \textbf{8}, 11 (2005).

\bibitem{Ash-view}A.~Ashtekar, {\it Loop quantum cosmology: An overview,} Gen. Rel. Grav. {\bf41}, 707 (2009).


\bibitem{AS11} A. Ashtekar, P. Singh, {\it Loop quantum cosmology: A status
report,}  Class. Quant. Grav. {\bf28}, 213001 (2011).

\bibitem{BCM} K. Banerjee, G. Calcagni, M. Mart¨ªn-Benito, {\it Introduction to Loop Quantum Cosmology,} SIGMA {\bf 8}, 016 (2012).


\bibitem{APS3} A. Ashtekar, T. Pawlowski, P. Singh,  {\it Quantum nature of the big bang: Improved
dynamics,}  Phys. Rev. D {\bf 74}, 084003 (2006).

\bibitem{ACS}A. Ashtekar, A. Corichi, and P. Singh, {\it Robustness of key features of loop quantum
cosmology,} Phys. Rev. D  {\bf 77}, 024046 (2008).

\bibitem{Zhang14}X. Zhang, {\it Loop quantum cosmology in 2+1 dimension,} Phys. Rev. D  {\bf 90}, 124018 (2014).

\bibitem{BTTa} N. Bodendorfer,  T. Thiemann and A. Thurn, {\it New variables for classical and quantum gravity in all dimensions I. Hamiltonian analysis}, Class. Quant. Grav, {\bf30}, 045001 (2013).

\bibitem{BTTb} N. Bodendorfer,  T. Thiemann and A. Thurn, {\it New variables for classical and quantum gravity in all dimensions II. Lagrangian analysis}, Class. Quant. Grav. {\bf30}, 045002 (2013).

\bibitem{BTTc} N. Bodendorfer,  T. Thiemann and A. Thurn, {\it New variables for classical and quantum gravity in all dimensions III. Quantum theory}, Class. Quant. Grav. {\bf30}, 045003 (2013).

\bibitem{BTTd} N. Bodendorfer,  T. Thiemann and A. Thurn, {\it New variables for classical and quantum gravity in all dimensions IV. Matter Coupling}, Class. Quant. Grav. {\bf30}, 045004 (2013).

\bibitem{Taveras}
V. Taveras, {\it Corrections to the Friedmann equations from loop quantum gravity for a universe with a free scalar field,} Phys. Rev. D \textbf{78}, 064072 (2008).

\bibitem{DMY}
Y. Ding, Y. Ma and J. Yang, {\it Effective scenario of loop quantum cosmology,} Phys. Rev. Lett. \textbf{102}, 051301
(2009).

\bibitem{YDM}
J. Yang, Y. Ding and Y. Ma, {\it Alternative quantization of the Hamiltonian in loop quantum cosmology,} Phys. Lett. B \textbf{682}, 1 (2009).

\bibitem{Boj11}
M. Bojowald, D. Brizuela, H. H. Hernandez, M. J. Koop, H. A.
Morales-Tecotl, {\it High-order quantum back-reaction and quantum cosmology with a positive cosmological constant,} Phys. Rev. D \textbf{84}, 043514 (2011).

\bibitem{ACH102}
A. Ashtekar, M. Campiglia, A. Henderson, {\it Loop quantum cosmology and spin foams,} Phys. Lett. B \textbf{681}, 347 (2009); {\it

Casting loop quantum cosmology in the spin foam paradigm,} Class. Quant. Grav. \textbf{27}, 135020 (2010); {\it Path integrals and the WKB approximation in loop quantum cosmolog,} Phys. Rev. D \textbf{82}, 124043 (2010).

\bibitem{QHM}
L. Qin, H. Huang and Y. Ma, \emph{Path integral and effective
Hamiltonian in loop quantum cosmology}, Gen. Rel.
Grav. {\bf 45}, 1191 (2013).

\bibitem{QDM}
L. Qin, G. Deng and Y. Ma, {\it Path integral and effective Hamiltonian in loop quantum cosmology,} Commun. Theor. Phys. \textbf{57}, 326
(2012).

\bibitem{QM1}
L. Qin and Y. Ma, {\it Coherent state functional integrals in quantum cosmology,} Phys. Rev. D \textbf{85}, 063515 (2012).

\bibitem{QM2}
L. Qin and Y. Ma, {\it Coherent state functional integral in loop quantum cosmology: Alternative dynamics,} Mod. Phys. Lett. \textbf{27}, 1250078 (2012).


\bibitem{AC13a}E. Alesci and F. Cianfrani, {\it A new perspective on cosmology in Loop Quantum Gravity,} Europhysics Lett.  {\bf 104}, 10001 (2013).

\bibitem{AC13b}E. Alesci and F. Cianfrani, {\it Quantum-reduced loop gravity: Cosmology,} Phys. Rev. D  {\bf 87}, 083521  (2013).

\bibitem{NB15}N. Bodendorfer, {\it A quantum reduction to Bianchi I models in loop quantum gravity ,} Phys. Rev. D {\bf91},  081502(R)
(2015).



\end{thebibliography}
\end{document}